\documentclass[10pt,twocolumn,twoside,]{IEEEtran}
\usepackage{ntheorem}
\usepackage{amsmath}
\usepackage{amsfonts}
\newtheorem{theorem}{Theorem}
\newtheorem{corollary}{Corollary}
\newtheorem{lemma}{Lemma}
\newtheorem{definition}{Definition}
\newtheorem{assumption}{Assumption}
\newtheorem{remark}{Remark}
\usepackage{ntheorem}
\usepackage{amsmath}
\usepackage{amsfonts}
\usepackage{listings}
\usepackage{color}
\usepackage{array}
\usepackage{tabu}
\usepackage{mathrsfs,amssymb}
\usepackage{graphicx}
\graphicspath{ {figure/} }
\usepackage{float}
\usepackage{algorithm}

{}

\usepackage{algpseudocode}
\algdef{SE}[DOWHILE]{Do}{doWhile}{\algorithmicdo}[1]{\algorithmicwhile\ #1}%
\DeclareMathOperator*{\argmin}{arg\,min}

\DeclareMathOperator*{\card}{card}
\DeclareMathOperator*{\supp}{supp}

\begin{document}
	
		\title{A Secure Sensor Fusion Framework for Connected and Automated Vehicles under Sensor Attacks}
	
	%
	%
	%

	\author{Tianci~Yang,
		~Chen~Lv
		\thanks{This work was supported by the SUG-NAP Grant (No. M4082268.050) of Nanyang Technological University, Singapore.}
		\thanks{The authors are with the School of Mechanical and Aerospace Engineering, Nanyang Technological University, Singapore.
			{\tt\small tianci.yang@ntu.edu.sg}}
	}
	%
	%

	\markboth{Journal of \LaTeX\ Class Files,~Vol.~14, No.~8, April~2020}%
	{Shell \MakeLowercase{\textit{et al.}}: Bare Demo of IEEEtran.cls for IEEE Journals}
	%



	\maketitle

	\begin{abstract}
	 By using various sensors to measure the surroundings and sharing local sensor information with the surrounding vehicles through wireless networks, connected and automated vehicles (CAVs) are expected to increase safety, efficiency, and capacity of our transportation systems. However, the increasing usage of sensors has also increased the vulnerability of CAVs to sensor faults and adversarial attacks. Anomalous sensor values resulting from malicious cyberattacks or faulty sensors may cause severe consequences or even fatalities. In this paper, we increase the resilience of CAVs to faults and attacks by using multiple sensors for measuring the same physical variable to create redundancy. We exploit this redundancy and propose a sensor fusion algorithm for providing a robust estimate of the correct sensor information with bounded errors independent of the attack signals, and for attack detection and isolation. The proposed sensor fusion framework is applicable to a large class of security-critical Cyber-Physical Systems (CPSs). To minimize the performance degradation resulting from the usage of estimation for control, we provide an $H_{\infty}$ controller for CACC-equipped CAVs capable of stabilizing the closed-loop dynamics of each vehicle in the platoon while reducing the joint effect of estimation errors and communication channel noise on the tracking performance and string behavior of the vehicle platoon. Numerical examples are presented to illustrate the effectiveness of our methods.
\end{abstract}

	\section{Introduction}
		Our transportation systems are enhancing its intelligence with the surge of connected and automated vehicles (CAVs). With the increasing connectivity and automation, CAVs are expected to provide a safer, more efficient and environmentally friendly transportation systems \cite{litman2017autonomous}\cite{anderson2014autonomous}. Compared with traditional ones, CAVs use a variety of sensors to measure the surroundings so that various driving tasks such as lane keeping, collision avoidance, and vehicle following can be fulfilled. Moreover, taking the advantage of wireless communications, CAVs are now able to share their local sensor information with the surrounding vehicles and roadside units to increase traffic efficiency, safety or form vehicle platoonings. Therefore, the accuracy of the on-board sensors has to be guaranteed to achieve safe and secure driving and anomalous sensor values which result from malicious cyberattacks or faulty sensors can lead to severe consequences or fatalities in the worst case \cite{petit2014potential}. However, sensors of CAVs might be designed without security considerations and hence remain vulnerable to adversarial attacks. For instance, in \cite{harris2015researcher}, a self-driving vehicle based on LiDAR was compromised remotely by an attacker who generates more echoes of objects and vehicles. In \cite{yan2016can}, the authors investigate the vulnerabilities of millimeter-wave radars, ultrasonic sensors and forward-looking cameras to attacks and show that off-the-shelf hardware can be used to perform jamming and spoofing attacks on a Tesla Model S automobile. It has been shown in \cite{petit2015remote} that inexpensive commodity hardware can be used to perform various types of attacks, e.g., blinding, jamming, replay and spoofing attacks on camera and LiDAR systems. This shows that we need strategic countermeasures for identifying and dealing with sensor attacks on CAVs. 
		
	Security and privacy problems for general cyber-physical systems under sensor attacks have been addressed using various methodologies \cite{Fawzi2014a}\nocite{Pasqualetti123}\nocite{Carlos_Justin4}\nocite{Chong2015}\nocite{Yang2018a} \nocite{Showkatbakhsh2017}\nocite{yang2020multi}-\cite{Tang2019}. Traditional works on autonomous vehicles security focus on the cryptography part, e.g., designing new protocols for device authentication and data transmission \cite{Chen2019,contreras2017internet}. There are only a few results addressing and solving the problem of quantifying (minimizing) the performance degradation induced by attacks on CAVs.
		 In \cite{Liu2019,Wyk2019}, the authors exploit sensor redundancy and provide detection and isolation algorithms for a single vehicle under sensor attacks. The problem of achieving consensus against replay attacks in an operator-vehicle network is solved in \cite{Elizabeth2012}. A control scheme is provided in \cite{Biron2017} for a vehicle platoon whose communication network, DSRC, is under Denial-of-Service (DoS) attack. In \cite{Merco2018}, suitable countermeasures to detect replay attack for connected vehicles are provided. The authors of \cite{mousavinejad2019distributed} provide an algorithm for detecting sensor attacks on connected vehicles using a set-membership filtering technique. The problem of attack detection and estimation for connected vehicles under sensor attacks is solved using an unbiased finite implulse response (UFIR) estimator in \cite{ju2020deception}. Shoukry et al. \cite{shoukry2015pycra} suggest randomly turning off the transmitter for detecting attacks on LiDARs, which however may degrade system performance. Attack-resilient sensor fusion algorithms for cyber-physical systems are proposed in \cite{ivanov2016attack}\cite{yang2018sensor}, where multiple so-called \textit{abstract} sensors are used to measure the same physical variable and some of them are under attacks. Each abstract sensor provides a set with all possible values for the true state of the variable for the fusion algorithm, and then an attack-free fused measurement is obtained by checking the intersections of these sets. We provide a secure sensor fusion framework for CAVs under sensor attacks, where noise on sensor measurements is assumed to be bounded with \textit{unknown} bounds, which might be applicable to a large class of security-critical CPSs. 		

Our sensor fusion approach is inspired by the work in \cite{Chong2015}, where the problem of state estimation for \emph{continuous-time linear time-invariant (LTI) systems} is addressed. The authors propose a way of constructing an estimator using a bank of Luenberger observers, that provides a robust state estimation despite the fact that sensor attacks are present. The main idea of their estimation scheme is to place redundant sensors in the system. Sensor \emph{redundancy} has been proved to be crucial for estimation under attacks \cite{Fawzi2014,Chong2015,Shoukry2017,Kim2016a,Tang2019,Mo2015}. Using redundant sensors/actuators for secure estimation and control is a commonly adopted technique, see, e.g., \cite{Fawzi2014,Chong2015,Shoukry2017,Kim2016a, Yang2018a, Yang2020}, and references therein. Note that it might be costly to create redundancy, which indicates some of these estimation methods might have conservative applications;
however, for security-critical systems, for instance, CAVs, this is the price to pay \cite{petit2015remote}.

In this manuscript, exploiting sensor redundancy, we address the problem of robust sensor fusion, attack detection and isolation, and control for a Cooperative Adaptive Cruise Control (CACC) equipped CAV in a vehicle platoon under sensor attacks. Assuming each vehicle in the platoon is equipped with multiple on-board sensors that measure the same physical variable, we consider the case where all the sensor measurements are subject to bounded noise with unknown bounds and the adversary has limited capability such that only less than half of the sensors are compromised. We assume that the set of attacked sensors are not a priori known to us and can be \textit{time-varying}. For instance, the time-variance of the set of attacked sensors may be caused by the attack strategy adopted by the attacker with energy-saving considerations \cite{zakerhaghighi2020implementation}. For every subset of sensors, we compute the largest difference between the average value of all the sensor measurements and the measurement of every single sensor in the subset. Then we select the sensors that lead to the smallest difference. If  some of the sensors are under attacks, their measurements produces larger difference than the attack-free ones, in general and hence will not be used by the algorithm for estimation. Our fusion algorithm provides estimation errors that are bounded independent of the attack signals. Once we have an attack-free measurement, we provide algorithms for detecting and isolating the attacked sensors under the condition that the noise bounds are known. Finally, for each vehicle in the vehicular platoon, we provide a robust CACC scheme that stabilizes its closed-loop dynamics while minimizing the joint effect of estimation errors and channel noise on the platooning string-stable behavior as well as the tracking performance of each vehicle. Our controller design method is an extension of the results given in \cite{Ploeg2014} by assuming system disturbances are present. In the case of cyberattacks, we show that a separation principle between sensor fusion and control holds and the vehicle platoon can be stabilized by closing the loop with the fusion algorithm and the robust controller. We conclude on stability of the closed-loop
dynamics of each vehicle in the platoon from Input-to-State Stability (ISS) \cite{Sontag2008} of the closed-loop system with respect to the bounded estimation error. The performance of our methods is illustrated in several numerical examples.

The paper is organized as follows. In Section \uppercase\expandafter{\romannumeral2}, some preliminary results needed for the subsequent sections are presented. In Section \uppercase\expandafter{\romannumeral3}, we describe the vehicle platoon system we consider. In Section \uppercase\expandafter{\romannumeral4}, we show that our fusion scheme provides robust estimates of the sensor measurements despite sensor attacks. In Section \uppercase\expandafter{\romannumeral5}, algorithms for detecting and isolating sensor attacks are presented. The proposed robust CACC scheme and stability analysis are given in Section \ref{control}. Simulation results are shown in Section \ref{sim}. Finally, in Section \ref{conclusion}, concluding remarks are given.
\section{Preliminaries}
\subsection{Notation}
We denote the set of real numbers by $\mathbb{R}$, the set of natural numbers by $\mathbb{N}$, and $\mathbb{R}^{n\times m}$ the set of $n\times m$ matrices for any $m,n \in \mathbb{N}$. For any vector $v\in\mathbb{R}^{n}$,  we denote {$v^{J}$} the stacking of all $v_{i}$, $i\in J$, $J\subset \left\lbrace 1,\hdots,n\right\rbrace$, $|v|=\sqrt{v^{\top} v}$, and $\supp(v)=\left\lbrace i\in\left\lbrace 1,\hdots,n\right\rbrace |v_{i}\neq0\right\rbrace $. For a sequence $\left\lbrace v(t)\right\rbrace _{t=0}^{\infty}$,  $||v||_{\infty} := \sup_{t\geq 0}|v(t)|$, $v(t) \in \mathbb{R}^{n}$. We say that a sequence $\left\lbrace v(t)\right\rbrace$ belongs to $l_{\infty}$, $\left\lbrace v(t)\right\rbrace \in l_{\infty}$, if $||v||_{\infty}<\infty$. $||v(t)||_{\mathcal{L}_{p}}$ is the p-norm of signal $v(t)$. We denote the cardinality of a set $S$ as $\card(S)$. The binomial coefficient is denoted as $\binom{a}{b}$, where $a,b$ are nonnegative integers. We denote a variable $m$ uniformly distributed in the interval $(z_{1},z_{2})$ as $m\sim\mathcal{U}(z_{1},z_{2})$ and normally distributed with mean $\mu$ and variance $\sigma^2$ as $m\sim \mathcal{N}(\mu,\sigma^2)$.
	\begin{definition}[Vehicle String Stability]\cite{Ploeg2011}\label{d1}
		Consider a string of $m\in N$ interconnected vehicles. This system is string stable
		if and only if
		\begin{equation}
			\begin{split}
				||z_{i}(t)||_{\mathcal{L}_{p}} \leq ||z_{i-1}(t)||_{\mathcal{L}_{p}},\hspace{2mm} \forall t\geq 0, 2 \leq i \leq m,	
			\end{split}
		\end{equation}
		where $z_{i}(t)$ can either be the distance error $e_{i}(t)$, the velocity
		$v_{i}(t)$ or the acceleration $a_{i}(t)$ of vehicle i; $z_{1}(t)\in\mathcal{L}_{p}$ is a
		given input signal, and $z_{i}(0) = 0$ for $2 \leq i \leq m$.
\end{definition}
	\section{System Description}
	\begin{figure}[t]\centering
		\includegraphics[width=0.5\textwidth]{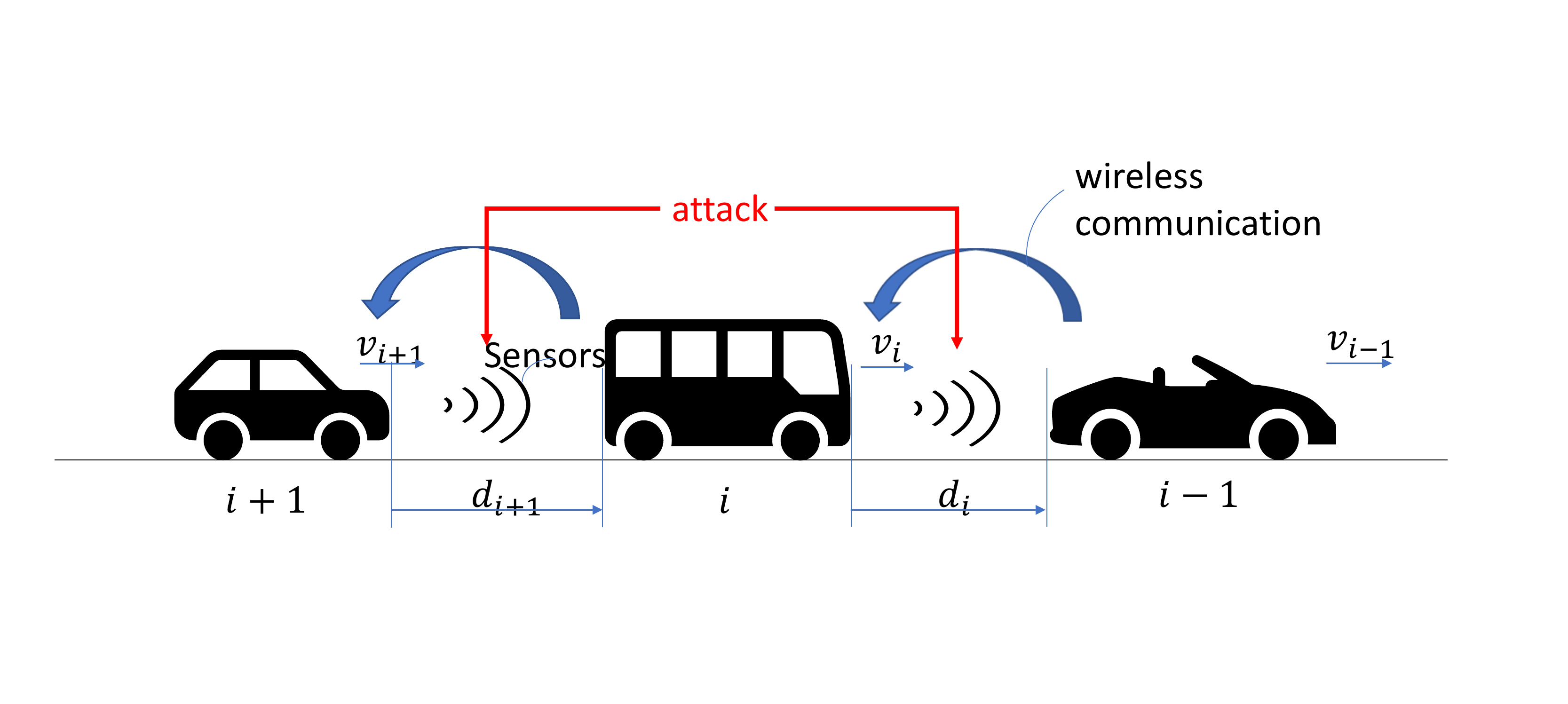}
		\caption{CACC-equipped vehicle platoon: each vehicle is equipped with multiple on-board sensors that measure the same physical variable while some are under attacks.}
		\centering
		\label{fig:1}
	\end{figure}
	Consider a platoon of $m$ vehicles, shown in Figure \ref{fig:1}.  Denote the distance between vehicle $i$ and its preceding vehicle $i-1$ as $d_{i}$, and its velocity as $v_{i}$. The objective of each vehicle is to keep a desired distance with its preceding vehicle:
	\begin{equation}
	d_{r,i}(t)=r_{i}+h_{i}v_{i}(t), i\in S_{m},
	\end{equation} 
	where $h_{i}$ represents the time headway of the $i$-th vehicle, $r_{i}$ is the standstill distance. $S_{m}=\left\lbrace i\in\mathbb{N}|1\leq i\leq m\right\rbrace $ denotes the set of all vehicles in a platoon of length $m\in\mathbb{N}$. The spacing policy adopted here is expected to improve string stability  \cite{Paper1996, Ploeg2014}. The spacing error $e_{i}(t)$ is then defined as 
	\begin{equation}
	\begin{split}
	e_{i}(t)=&d_{i}(t)-d_{r,i}(t),\\
	=&(q_{i-1}(t)-q_{i}(t)-L_{i})-(r_{i}+h_{i}v_{i}(t)),
	\end{split}
	\end{equation}
	where $q_{i}$ denotes the rear-bumper position of vehicle $i$ and $L_{i}$ denotes its length. We consider the following vehicle model adopted in \cite{ploeg2011design}, 
	\begin{equation}
		\begin{split}
		\begin{bmatrix}
		\dot{d}_{i}\\
		\dot{v}_{i}\\
		\dot{a}_{i}
		\end{bmatrix}=\begin{bmatrix}
		v_{i-1}-v_{i}\\
		a_{i}\\
		-\frac{1}{\tau_{i}}a_{i}+\frac{1}{\tau_{i}}u_{i}
		\end{bmatrix},\hspace{2mm}i\in S_{m},
		\end{split}
	\end{equation}
	with $\tau_{i}$ being a time constant representing
	driveline dynamics of vehicle $i$.  $a_{i}$ is the acceleration of vehicle $i$; $u_{i}$ is the desired accelerations of vehicles $i$. We adopt the controller described in \cite{ploeg2011design}, where a new input $\xi_{i}$ is defined such that
	\begin{equation}
		\begin{split}
		h_{i}\dot{u}_{i}=-u_{i}+\xi_{i},
		\end{split}
	\end{equation}
	with $\xi_{i}$ given as follows:
	\begin{equation}
		\xi_{i}=K\begin{bmatrix}
		e_{i}\\
		\dot{e}_{i}\\
		\ddot{e}_{i}
		\end{bmatrix}+u_{i-1}=K\begin{bmatrix}
		d_{i}-h_{i}v_{i}\\
		v_{i-1}-v_{i}-h_{i}a_{i}\\
		a_{i-1}-a_{i}-h_{i}\dot{a}_{i}
		\end{bmatrix}+u_{i-1},\hspace{1mm}i\in S_{m}.
	\end{equation}
	
The feedforward term $u_{i-1}$ is transmitted from vehicle $i-1$ to vehicle $i$ through wireless communication channels. $v_{i}$ is measured by a wheel speed sensor of vehicle $i$, $a_{i}$ is measured by its accelerator. $d_{i}$ is measured by e.g, an ultrasonic sensor, a LiDAR, a radar, or a vision sensor. 
\section{Secure Sensor Fusion under Sensor Attacks}\label{estimation}
Compared with the wheel speed sensor and the accelerator, which might require attackers' physical access to be compromised, these sensors that measure $d_{i}$ are more exposed to the adversarial environment as they can be attacked from a distance \cite{cao2019adversarial,bannasch2012method,petit2015remote,yan2016can}. Therefore, we assume each vehicle in the platoon is equipped with $N$ different on-board sensors for measuring $d_{i}$, i.e., at every time $t$, vehicle $i$ has $N$ different measurements of $d_{i}(t)$. These measurements can be provided by a combination of various types of sensors, e.g., a combination of an ultrasonic sensor, a milimeter wave radar and a LiDAR, or multiple same type of sensors, e.g., multiple radars, or multiple LiDARs with different wavelengths etc.. Then, we assume $q$ ($q<N$) measurements are under attacks:
\begin{equation}\label{sss}
\begin{split}
D(t)=\left[ \begin{matrix}
d_{i}(t)+\nu_{1}(t)\\
d_{i}(t)+\nu_{2}(t)\\
\vdots\\
d_{i}(t)+\nu_{N}(t)
\end{matrix}\right] +\eta(t)=\left[ \begin{matrix}
D_{1}(t)\\
D_{2}(t)\\
\vdots\\
D_{N}(t)
\end{matrix}\right]
\end{split}
\end{equation}
where $\nu_{i}\in\mathbb{R}$ and $\left\lbrace \nu_{i}(t)\right\rbrace \in l_{\infty}$ denotes measurement noise in each sensor, and $\eta\in\mathbb{R}^{N}$ denotes the vector of attack, i.e., if the $i$-th sensor is compromised, then $\eta_{i}(t)\neq 0$ for some $t\geq 0$; otherwise $\eta_{i}(t)=0$ for all $t\geq 0$. Let $W(t)\subset\left\lbrace 1,\ldots,N\right\rbrace $ represent the unknown set of attacked sensors at time $t$, i.e.,
\begin{equation}
\supp(\eta(t))\subseteq W(t).
\end{equation}
We assume an attacker attacks a sensor by modifying its measurement to any arbitrary value.
\begin{remark}
	It is usually assumed in the literature, e.g., \cite{Chong2015,Fawzi2014a,Showkatbakhsh2017,Yang2018a,Shoukry2017,Kim2016a,Yang2020} that the set of attacked nodes is time-invariant, i.e., the attacker cannot change his mind by attacking the other nodes rather than the ones he chose initially. We relax this assumption by allowing the set of attacked sensors to be time-varying, i.e., the attacker can choose a different set of sensors to compromise at different time.
\end{remark}
\begin{definition}[Reconstructability]
	$d_{i}(t)$ is reconstructible from $D(t)$ under $q$ attacks for all $t\geq 0$, if for every $d_{i}(t)$, $\bar{d}_{i}(t)$, sets $W_{a}(t)$, $W_{b}(t)\subset\left\lbrace 1,\ldots, N\right\rbrace $ with $\card(W_{a}(t))$, $\card(W_{b}(t))\leq q$, and $\supp(\eta_{a}(t))\subseteq W_{a}(t)$, $\supp(\eta_{b}(t))\subseteq W_{b}(t) $, and all $t\geq 0$ we have
	\begin{equation}
	D(t)=\bar{D}(t)\Longrightarrow d_{i}(t)=\bar{d}_{i}(t).
	\end{equation}
\end{definition}

This definition means that if $d_{i}(t)$ is reconstructible from $D(t)$ under $q$ attacks, then there do not exist two distinct values $d_{i}(t)$ and $\bar{d}_{i}(t)$ that can explain the received data with less than $q$ sensors attacked, hence $d_{i}(t)$ can be reconstructed unambiguously.
\begin{theorem}\label{b}
	$d_{i}(t)$ is reconstructible from $D(t)$ under $q$ attacks for all $t\geq 0$ if and only if $q<\frac{N}{2}$.
\end{theorem}
\textit{Proof:}

\textbf{1) if:}  Suppose that $q<\frac{N}{2}$.  For all $\eta_{a}(t)$ and $\eta_{b}(t)$ with $\card(\supp(\eta_{a}(t)))\leq q$ and $\card(\supp(\eta_{b})(t))\leq q$ and $t\geq 0$, the nonzero elements of $\eta_{b}(t)-\eta_{a}(t)$ cannot exceed $2q$, which is strictly smaller than $N$. Therefore, there do not exist $\eta_{a}(t)\neq \eta_{b}(t)$ such that
\begin{equation}\label{e}
	\eta_{b}(t)-\eta_{a}(t)=\left[ \begin{matrix}
		d_{i}(t)-\bar{d}_{i}(t)\\
		d_{i}(t)-\bar{d}_{i}(t)\\
		\vdots\\
		d_{i}(t)-\bar{d}_{i}(t)
	\end{matrix}\right],
\end{equation}
for any $t\geq 0$, since the number of nonzero elements on the right side of \eqref{e} is equal to $N$. Therefore, there do not exist $d_{i}(t)\neq\bar{d}_{i}(t)$ such that
\begin{equation}\label{eq1}
	\left[ \begin{matrix}
		d_{i}(t)+m_{1}(t)\\
		d_{i}(t)+m_{2}(t)\\
		\vdots\\
		d_{i}(k)+m_{N}(t)
	\end{matrix}\right] +\eta_{a}(t)=\left[ \begin{matrix}
		\bar{d}_{i}(t)+m_{1}(t)\\
		\bar{d}_{i}(t)+m_{2}(t)\\
		\vdots\\
		\bar{d}_{i}(t)+m_{N}(t)
	\end{matrix}\right] +\eta_{b}(t),
\end{equation}
for any $t\geq 0$, which indicates $d_{i}(t)$ is reconstructible from $D(t)$ for all $t\geq 0$.

\textbf{2) only if:}
We prove by contrapositive. Suppose $q\geq\frac{N}{2}$, then for all $W_{a}(t)\subset\left\lbrace 1, \ldots, N\right\rbrace $, with $\card(W_{a}(t))=q$ and $t\geq 0$, there must exist another set $W_{b}(t)\subset\left\lbrace 1, \ldots, N\right\rbrace $ with $\card(W_{b}(t))=q$ such that $W_{b}(t)\cup W_{a}(t)=\left\lbrace 1, \ldots, N\right\rbrace $ since $2q\geq N$. Let $I_{ab}(t)=W_{a}(t)\cap W_{b}(t)$. For all $d_{i}(t)\neq\bar{d}_{i}(t)$ and $t\geq 0$, we choose 
\begin{equation}
	\begin{split}
		&\eta_{bi}(t)=d_{i}(t)-\bar{d}_{i}(t), \eta_{ai}(t)=0, i\in \left\lbrace W_{b}(t)\setminus I_{ab}(t)\right\rbrace \\
		&\eta_{bi}(t)=0, \eta_{ai}(t)=\bar{d}_{i}(t)-d_{i}(t), i\in \left\lbrace W_{a}(t)\setminus I_{ab}(t)\right\rbrace \\
		&\eta_{bi}(t)=d_{i}(t),\eta_{ai}(t)=\bar{d}_{i}(t), \hspace{6mm}i\in I_{ab}(t),
	\end{split}
\end{equation}
then \eqref{eq1} holds with $d_{i}(t)\neq\bar{d}_{i}(t)$ for all $t\geq 0$, which indicates
\begin{equation}
	D(t)=\bar{D}(t),
\end{equation}
with $d_{i}(t)\neq \bar{d}_{i}(t)$ for all $t\geq 0$. This contradicts the statement that $d_{i}(t)$ is reconstructible from $D(t)$ under $q$ attacks for all $t\geq 0$.\hfill$\blacksquare$. 
\begin{assumption}\label{a1}
	The number of attacked sensors does not exceed $\frac{N}{2}$, i.e.,
	\begin{equation}
	\card(W(t))\leq q<\frac{N}{2}.
	\end{equation}
\end{assumption}
\begin{remark}
	Assumption \ref{a1} allows that all sensors are compromised at some time $t\geq 0$, but at each time, attackers only inject attack signals into less than $q$ sensors. This limitation may be caused by energy or hardware constraints, or attack strategy adopted. For instance, an attacker may use a laser for compromising all the vision sensors at different time, but at each time, only one single vision sensor may be affected due to a small beam length of the laser \cite{petit2015remote}; a radio generator capable of sending radio signals with different frequencies can be used to compromise multiple radars at different time, but only one radar can be affected at each time \cite{zakerhaghighi2020implementation}.  
\end{remark}
\begin{corollary}\label{c2}
	Under Assumption \ref{a1}, among all $N$ sensors that measure $d_{i}$, at least $N-q$ of them are attack-free; among every set of $N-q$ sensors that measure $d_{i}$, at least $N-2q$ of them are attack-free
\end{corollary}
For every subset $J\subset\left\lbrace 1, \ldots, N\right\rbrace $ sensors and $t\geq 0$, we define $\hat{d}_{J}(t)$ as the average value of all the measurements given by subset $J$ of sensors at time $t$, as follows:
\begin{equation}
\hat{d}_{J}(t)\triangleq  \frac{\sum_{i\in J}D_{i}(t)}{\card(J)}
\end{equation}
Define
\begin{equation}
||\nu||_{\infty}\triangleq\max_{i\in\left\lbrace 1,\cdots,N\right\rbrace }\left\lbrace ||\nu_{i}||_{\infty}\right\rbrace.
\end{equation} 
\begin{remark}
	We assume that $||\nu_{i}||_{\infty}$ and $||\nu||_{\infty}$ are both unknown.
\end{remark}
\begin{lemma}
	If $\eta^{J}(t)=0$, then
	\begin{equation}
	|\hat{d}_{J}(t)-d_{i}(t)|\leq ||\nu||_{\infty}.
	\end{equation}
	for all $t\geq 0$.
\end{lemma}
\textit{Proof:}
Since
\begin{equation}
\begin{split}
\left|\sum_{i}^{n}\nu_{i}\right|=&\sqrt{(\nu_{1}+\nu_{2}+\cdots+\nu_{n})^{2}}\\
=&\sqrt{\nu_{1}^{2}+\nu_{2}^{2}+\cdots+\nu_{n}^{2}+2(\nu_{1}\nu_{2}+\cdots+\nu_{n-1}\nu_{n})}\\
\leq&\sqrt{\nu_{1}^{2}+\nu_{2}^{2}+\cdots+\nu_{n}^{2}+(\nu_{1}^{2}+\nu_{2}^{2}+\cdots+\nu_{n-1}^{2}+\nu_{n}^{2})}\\
=&\sqrt{\left( 1+\left( \begin{matrix}
	1\\
	n-1
	\end{matrix}\right) \right) \left( \nu_{1}^{2}+\nu_{2}^{2}+\cdots+\nu_{n}^{2}\right\rbrace }\\
\leq&\sqrt{n^{2}||\nu||_{\infty}^{2}}\\
=&n||\nu||_{\infty},
\end{split}
\end{equation}
we have
\begin{equation}
\begin{split}
|\hat{d}_{J}(t)-d_{i}(t)|=&\left| \frac{\sum_{i\in J}D_{i}(t)}{\card(J)}-d_{i}(t)\right|\\
=&\frac{1}{\card(J)}\left|\sum_{i\in J}\nu_{i}\right|\\
\leq&\frac{1}{\card(J)}\card(J)||\nu||_{\infty}\\
=&||\nu||_{\infty}.
\end{split}
\end{equation}
\hfill$\blacksquare$

Under Assumption \ref{a1}, there exists at least one subset $\bar{I}(t)\subset \left\lbrace 1, \ldots, N\right\rbrace $ with $\card(\bar{I}(t))=N-q$ such that $\eta^{\bar{I}(t)}(t)=0$ for $t\geq 0$. Then, the difference between $\hat{d}_{\bar{I}(t)}(t)$ and any $D_{i}(t)$, $i\in\bar{I}(t)$ will be smaller than the other subsets $J\subset\left\lbrace 1, \ldots, N\right\rbrace $ with $\card(J)=N-q$ and $\eta^{J}(t)\neq 0$, in general. This motivates the following fusion algorithm.

For every subset $J\subset\left\lbrace 1,\ldots,N\right\rbrace $ of sensors with $\card(J)=N-q$, define $\pi_{J}(t)$ as the largest difference between $\hat{d}_{J}(t)$ and $D_{i}(t)$ for all $i\in J$, i.e.,
\begin{equation}\label{es1}
\pi_{J}(t)=\underset{i\in J}{\max}\left| \hat{d}_{J}(t)-D_{i}(t)\right| 
\end{equation}
for all $t\geq 0$, and the sequence $\sigma(t)$ is given as,
\begin{equation}\label{es2}
\begin{split}
\sigma(t)=\underset{J\subset\left\lbrace 1,\ldots, N\right\rbrace: \card(J)=N-q }{\argmin}\pi_{J}(t).
\end{split}
\end{equation}
Then, as proved below, the fused measurement indexed by $\sigma(t)$:
\begin{equation}\label{es3}
\hat{d}_{i}(t)=\hat{d}_{\sigma(t)}(t),
\end{equation}
is an attack-free measurement of $d_{i}(t)$. The following result uses the terminology presented above.
\begin{theorem}
	Consider the fusion algorithm \eqref{es1}-\eqref{es3}. Define the fused measurement error $e(t):=\hat{d}_{\sigma(t)}(t)-d_{i}(t)$, and let Assumption \ref{a1} be satisfied; then,
	\begin{equation}
	|e(t)|\leq 3||\nu||_{\infty}\label{sa},
	\end{equation}
	for all $t\geq 0$.
\end{theorem}
\textit{Proof:}
Under Assumption \ref{a1}, there exists at least one subset $\bar{I}(t)\subset\left\lbrace 1, \ldots, N\right\rbrace $ with $\card(\bar{I}(t))=N-q$ such that $\eta^{\bar{I}(t)}(t)=0$ for $t\geq 0$. Then 
\begin{equation}
|\hat{d}_{\bar{I}(t)}(t)-d_{i}(t)|\leq ||\nu||_{\infty}.
\end{equation}
Moreover, for all $i\in\bar{I}(t)$, $\eta_{i}(t)=0$ and we have
\begin{equation}
|D_{i}(t)-d_{i}(t)|=|\nu_{i}(t)|
\end{equation}
Then we have
\begin{equation}
\begin{split}
\pi_{\bar{I}}(t)=&\underset{i\in \bar{I}}{\max}\left| \hat{d}_{\bar{I}(t)}(t)-D_{i}(t)\right| \\
=&\underset{i\in \bar{I}}{\max}\left|\hat{d}_{\bar{I}(t)}(t)-d_{i}(t)+d_{i}(t)-D_{i}(t)\right| \\
=&\left| \hat{d}_{\bar{I}(t)}(t)-d_{i}(t)\right| +\underset{i\in \bar{I}}{\max}\left| d_{i}(t)-D_{i}(t)\right| \\
\leq&||\nu||_{\infty}+\underset{i\in \bar{I}}{\max}|\nu_{i}(t)|
\end{split}
\end{equation}
From Corollary \ref{c2}, among every set of $N-q$ sensors, at least one of the sensors is attack free since $N-2q\geq 1$. Therefore, there exists at least one $\bar{i}(t)\in\sigma(t)$ such that $\eta_{\bar{i}(t)}(t)=0$ for $t\geq 0$ and
\begin{equation}
\begin{split}
|D_{\bar{i}}(t)-d_{i}(t)|=|\nu_{\bar{i}(t)}(t)|
\end{split}
\end{equation}
From \eqref{es2}, we have $\pi_{\sigma(t)}(t)\leq\pi_{\bar{I}(t)}(t)$. From \eqref{es1}, we have
\begin{equation}
\begin{split}
\pi_{\sigma(t)}(t)=&\underset{i\in\sigma(t)}{\max}\left| \hat{d}_{\sigma(t)}(t)-D_{i}(t)\right| \\
\geq&\left| \hat{d}_{\sigma(t)}(t)-D_{\bar{i}(t)}(t)\right|
\end{split}
\end{equation}
Using the lower bound on $\pi_{\sigma(t)}(t)$ and the triangle inequality, we have that
\begin{equation}\label{ee}
\begin{split}
|e_{\sigma(t)}(t)|=&|\hat{d}_{\sigma(t)}-d_{i}(t)|\\
=&\left| \hat{d}_{\sigma(t)}(t)-D_{\bar{i}(t)}(t)+D_{\bar{i}(t)}(t)- d_{i}(t)\right| \\
\leq&\pi_{\sigma(t)}(t)+|\nu_{\bar{i}}(t)|\\
\leq&\pi_{\bar{I}}(t)+|\nu_{\bar{i}}(t)|\\
\leq&||\nu||_{\infty}+\underset{i\in \bar{I}}{\max}|\nu_{i}(t)|+|\nu_{\bar{i}}(t)|\\
\leq&3 ||\nu||_{\infty}
\end{split}
\end{equation}
Inequality \eqref{ee} is of the form \eqref{sa} and the result follows.\hfill$\blacksquare$\\[2mm]
\begin{remark}
Note that this secure sensor fusion scheme is general enough to be applicable to a large class of CPSs. For security-critical CPSs, sensor redundancy can be created for security purposes and then our sensor fusion framework can be applied to provide robust sensor information in spite of attacks.
\end{remark}
\begin{section}{Detection and isolation}
	In this section, we now assume the bounds on each sensor measurement noise is known to us, i.e., $||\nu_{i}||_{\infty} $ for all $i\in\left\lbrace 1, \ldots, N\right\rbrace $ is known. We first provide a simple technique for detecting sensor attacks. Then, we use the fusion algorithm presented in Section \ref{estimation} to select the attack-free sensors and isolate the ones that are compromised.
	\subsection{Detection strategy}
	If all the sensors are attack-free at time $t$, the deviation between $\frac{\sum_{i}^{N}D_{i}(t)}{N}$ and $D_{i}(t)$ for all $i\in\left\lbrace 1,\ldots,N\right\rbrace $ will be small and 
	\begin{equation}\label{d11}
	\begin{split}
	\left| \frac{\sum_{i=1}^{N}D_{i}(t)}{N}-D_{i}(t)\right|\leq&\left| \frac{\sum_{i=1}^{N}D_{i}(t)}{N}-d_{i}(t)\right|+|\nu_{i}(t)
	|\\ \leq&||\nu||_{\infty}+||\nu_{i}||_{\infty}, \forall i\in\left\lbrace 1,\ldots,N\right\rbrace
	\end{split} 
	\end{equation}
	Define
	\begin{equation}\label{d12}
	\tau_{di}\triangleq||\nu||_{\infty}+||\nu_{i}||_{\infty},
	\end{equation} 
	for $i\in\left\lbrace 1,\ldots, N\right\rbrace $. Then, attacks are detected at time $t$ if there exists one or more than one sensor $i\in\left\lbrace 1, \ldots, N\right\rbrace $ such that 
	\begin{equation}\label{d13}
	\left| \frac{\sum_{i=1}^{N}D_{i}(t)}{N}-D_{i}(t)\right|>\tau_{di},
	\end{equation}
	for some $t\geq 0$. However, note that it is still possible that $\eta(t)\neq 0$ for some $t\geq 0$ but inequality (\ref{d11}) still holds, which results in a failure of detection. To increase the detection rate of our algorithm, we perform attack detection over every time window of length $T \in \mathbb{N}$. That is, for each $t \in [iT,(i+1)T)$, $i \in \mathbb{N}$, we check if \eqref{d13} is satisfied for all $t$ in the time window. If there exists  $t_{1} \in [iT,(i+1)T)$, $i \in \mathbb{N}$ such that \eqref{d13} holds, we say that sensors are under attack in the $i$-th time window. Otherwise, we say sensors are attack-free in this time window. This detection procedure is formally stated in Algorithm \ref{alg:the_alg2}.
	
	\begin{algorithm}
		\caption{Attack Detection.}
		\label{alg:the_alg2}
		\begin{algorithmic}[1]
			\State \text{Fix the window size $T\in\mathbb{N}$}.
			\State For $ i\in\mathbb{N}$, if $\exists t_{1}\in [ iT, (i+1)T) $ such that \eqref{d13} holds,
				then sensor attacks occur in the $i$-th window, and
			\begin{equation*}
				detection(i)=1;
			\end{equation*}
			\text{otherwise, sensors are attack-free in the $i$-th window,} \text{and}
			\begin{equation*}
				detection(i)=0.
			\end{equation*}
			\State \text{Return $detection(i)$}
		\end{algorithmic}
	\end{algorithm}
	\subsection{Isolation strategy}
	From Section \ref{estimation}, we know that $\sigma(t)\subset \left\lbrace 1, \ldots, N\right\rbrace $ is a set of attack-free sensors. For $t\geq 0$, we randomly select one sensor $i^{*}(t)\in\sigma(t)$ and we have $\eta_{i^{*}(t)}(t)=0$. For each $i\in\left\lbrace 1, \ldots, N\right\rbrace $ and $t\geq 0$, we compute the difference between $D_{i}(t)$ and $D_{i^{*}(t)}(t)$. Then, if the $i$-th sensor is attack-free, i.e., $\eta_{i}(t)=0$, we have
	\begin{equation}\label{ee1}
	\begin{split}
	|D_{i^{*}(t)}(t)-D_{i}(t)|=&|D_{i^{*}(t)}(t)-d_{i}(t)+d_{i}(t)-D_{i}(t)|\\
	\leq&|D_{i^{*}(t)}(t)-d_{i}(t)|+|d_{i}(t)-D_{i}(t)|\\
	\leq&||\nu_{i^{*}(t)}||_{\infty}+||\nu_{i}||_{\infty}.
	\end{split}
	\end{equation}
	For each $i\in\left\lbrace 1, \ldots, N\right\rbrace $, define
	\begin{equation}\label{i1}
	\begin{split}
	\tau_{i}(t)\triangleq ||\nu_{i^{*}(t)}||_{\infty}+||\nu_{i}||_{\infty}.
	\end{split}
	\end{equation}
	Then for $t\geq 0$, the $i$-th sensor is isolated as an attacked one if
	\begin{equation}\label{i2}
	|D_{i^{*}(t)}(t)-D_{i}(t)|>\tau_{i}(t).
	\end{equation} 
	Then, the set of sensors that are isolated as the attacked ones at time $t$, which we denote as $\hat{W}(t)$, is given as 
	\begin{equation}\label{i3}
	\begin{split}
	\hat{W}(t)=\left\lbrace i\in\left\lbrace 1, \ldots, N\right\rbrace \left|  |D_{i^{*}(t)}(t)-D_{i}(t)|>\tau_{i}(t)\right.\right\rbrace .
	\end{split}
	\end{equation}
Thus, the set $\left\lbrace 1,\ldots,p\right\rbrace \setminus\hat{W}(t)$ is the set of attack-free sensors at time $t$.
\begin{figure}[t]\centering
	\includegraphics[width=0.5\textwidth]{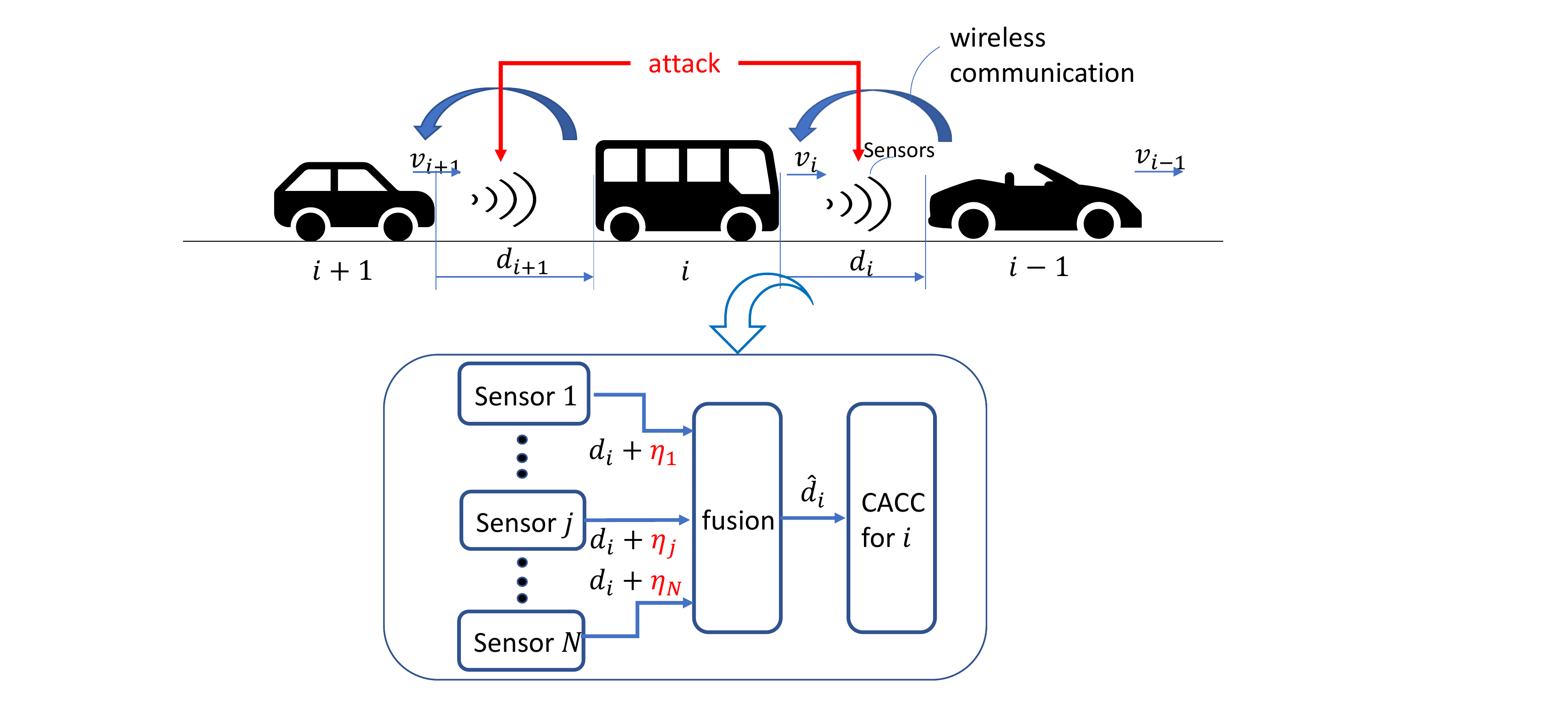}
	\caption{Schematic of the sensor fusion algorithm and CACC for vehicle $i$ under sensor attacks.}
	\centering
	\label{fig:sab}
\end{figure}
	\section{Control}\label{control}
We have shown that the error of the proposed fusion algorithm is bounded independent of attacks. In this section, we assume that the on-board sensors of each vehicle in the platoon and the inter-vehicle communication channels are noisy and each vehicle uses the estimation provided by the fusion algorithm for control, as shown in Figure \ref{fig:sab}. The estimation errors of the fusion algorithm introduce uncertainty into the system loop and may deteriorate the performance of the platooning. For each vehicle in the platoon, we consider designing an $H_{\infty}$ CACC scheme to stabilize its closed-loop dynamics while minimizing the joint effect of communication channel noise, sensor noise and estimation errors on platooning string stability and tracking performance of each vehicle.
Adopting the controller described in \cite{Ploeg2014} that fulfills the vehicle-following objective and the string stability, we consider a homogeneous platoon model given in \cite{Ploeg2014} with $h_{i}=h$, $\tau_{i}=\tau$ for all $i\in S_{m}$, and assume that the on-board sensors of vehicle $i$ that measure relative velocity, relative acceleration are perturbed by noise $\omega_{vi}$ and $\omega_{ai}$ respectively, $u_{i-1}$ is perturbed by channel noise $\omega_{ui}$, relative distance information is provided by the fusion algorithm and hence perturbed by $e_{\sigma(t)}$, which is formulated as follows:
\begin{equation}\label{e1}
	\begin{split}
		\begin{bmatrix}
			\dot{e}_{i}\\
			\dot{v}_{i}\\
			\dot{a}_{i}\\
			\dot{u}_{i}
		\end{bmatrix}=&\begin{bmatrix}
			0&-1&-h&0\\
			0&0&1&0\\
			0&0&-\frac{1}{\tau}&\frac{1}{\tau}\\
			\frac{k_{p}}{h}&-\frac{k_{d}}{h}&-k_{d}-\frac{k_{dd}(h-\tau)}{h\tau}&-\frac{k_{dd}h+\tau}{h\tau}
		\end{bmatrix}\begin{bmatrix}
			{e}_{i}\\
			{v}_{i}\\
			{a}_{i}\\
			{u}_{i}
		\end{bmatrix}\\
		&+\begin{bmatrix}
			0&1&0&0\\
			0&0&0&0\\
			0&0&0&0\\
			\frac{k_{p}}{h}&\frac{k_{d}}{h}&\frac{k_{dd}}{h}&\frac{1}{h}
		\end{bmatrix}\begin{bmatrix}
			e_{\sigma(t)}\\
			v_{i-1}+\omega_{vi}\\
			a_{i-1}+\omega_{ai}\\
			u_{i-1}+\omega_{ui}
		\end{bmatrix}, \hspace{3mm} i\in S_{m}.
	\end{split}
\end{equation}
The first vehicle, without a preceding vehicle in front, will follow a virtual reference vehicle $(i=0)$, so that the same controller as the other vehicles can be employed to the lead vehicle. We formulate the virtual reference vehicle as follows:
\begin{equation}\label{e2}
	\begin{split}
		\begin{bmatrix}
			\dot{e}_{0}\\
			\dot{v}_{0}\\
			\dot{a}_{0}\\
			\dot{u}_{0}
		\end{bmatrix}=\begin{bmatrix}
			0&0&0&0\\0&0&1&0\\0&0&-\frac{1}{\tau}&\frac{1}{\tau}\\0&0&0&-\frac{1}{h}
		\end{bmatrix}\begin{bmatrix}
			e_{0}\\
			v_{0}\\
			a_{0}\\
			u_{0}
		\end{bmatrix}+\begin{bmatrix}
			0\\0\\0\\-\frac{1}{h}
		\end{bmatrix}\xi_{0},
	\end{split}
\end{equation}
where $\xi_{0}$ denotes external platoon input.
Let $\omega_{i}=\begin{bmatrix}
	e_{\sigma(t)},v_{i-1}+                                        \omega_{vi},a_{i-1}+\omega_{ai},u_{i-1}+\omega_{ui}
\end{bmatrix}^{\top}$, $x_{i}=\begin{bmatrix}
	e_{i}&
	v_{i}&
	a_{i}&
	u_{i}
\end{bmatrix}^{\top}$. To implement the design method provided in \cite{Ploeg2014}, we formulate system \eqref{e1} in the following way:
\begin{equation}\label{ss}
	\left\{\begin{split}
		\dot{x}_{i}=&Ax_{i}+B_{1}w_{i}+B_{2}u_{i}\\
		\tilde{z}_{i}=&C_{1}x_{i}\\
		y_{i}=&C_{2}x_{i}+D_{21}w_{i}\\
		u_{i}=&Ky_{i}
	\end{split}\right.
\end{equation}
with 
\begin{equation}
	\begin{split}
		A=&\begin{bmatrix}
			0&-1&h&0\\
			0&0&0&1\\
			0&0&-\frac{1}{\tau}&\frac{1}{\tau}\\
			0&0&\frac{1}{h}&-\frac{1}{h}
		\end{bmatrix},C_{2}=\begin{bmatrix}
			1&0&0&0\\
			0&-1&-h&0\\
			0&0&-\frac{h}{\tau}&-\frac{h}{\tau}
		\end{bmatrix},\\
		B_{1}=&\begin{bmatrix}
			0&1&0&0\\
			0&0&0&0\\
			0&0&0&0\\
			0&0&0&\frac{1}{h}
		\end{bmatrix},B_{2}=\begin{bmatrix}
			0\\0\\0\\\frac{1}{h}
		\end{bmatrix},
		K=\begin{bmatrix}
			k_{p}&k_{d}&k_{dd}
		\end{bmatrix},\\
		D_{21}=&\begin{bmatrix}
			1&0&0&0\\
			0&1&0&0\\
			0&0&1&0
		\end{bmatrix},
	\end{split}
\end{equation}
where $\tilde{z}_{i}$ represents a vector of performance variables.
We are interested in the propagation of $v_{i}$ along the vehicle string, i.e., $z_{i}=v_{i}$ in Definition \ref{d1}. Hence,
to minimize the effect of $\omega_{i}$ on string stability and tracking performance, we let $C_{1}$ in \eqref{ss} be given as
\begin{equation}
	C_{1}=\begin{bmatrix}
		1&0&0&0\\
		0&1&0&0
	\end{bmatrix}.
\end{equation}
Then, a static-output feedback $H_{\infty}$ controller can be obtained by running the algorithms 3 and 4 in \cite{He2006} while letting $k_{p}$, $k_{d}>0$, $k_{dd}>0$ and $k_{d}>k_{p}\tau$ to fulfill the vehicle following control objective \cite{Ploeg2014}.

To provide a stability analysis of the closed-loop dynamics, system \eqref{e1} is  formulated as:
\begin{equation}\label{ee2}
	\dot{x}_{i}=A_{ci}x_{i}+\tilde{B}_{i}\omega_{i},\hspace{3mm}i\in S_{m}
\end{equation}
with
\begin{equation}
	\begin{split}
		A_{ci}=&\begin{bmatrix}
			0&-1&-h&0\\
			0&0&1&0\\
			0&0&-\frac{1}{\tau}&\frac{1}{\tau}\\
			\frac{k_{p}}{h}&-\frac{k_{d}}{h}&-k_{d}-\frac{k_{dd}(h-\tau)}{h\tau}&-\frac{k_{dd}h+\tau}{h\tau}
		\end{bmatrix},\\
		\tilde{B}_{i}=&\begin{bmatrix}
			0&1&0&0\\
			0&0&0&0\\
			0&0&0&0\\
			\frac{k_{p}}{h}&\frac{k_{d}}{h}&\frac{k_{dd}}{h}&\frac{1}{h}
		\end{bmatrix}.
	\end{split}
\end{equation}
Let $x_{0}=\begin{bmatrix}
	e_{0}&
	v_{0}&
	a_{0}&
	u_{0}
\end{bmatrix}^{\top}$, system \eqref{e2} can be formulated as
\begin{equation}
	\begin{split}
		\dot{x}_{0}=A_{c0}x_{0}+\tilde{B}_{0}\xi_{0},
	\end{split}
\end{equation}
with 
\begin{equation}
	\begin{split}
		A_{c0}=
		\begin{bmatrix}
			0&0&0&0\\0&0&1&0\\0&0&-\frac{1}{\tau}&\frac{1}{\tau}\\0&0&0&-\frac{1}{h}
		\end{bmatrix},
		\tilde{B}_{0}=\begin{bmatrix}
			0\\0\\0\\-\frac{1}{h}
		\end{bmatrix}.
	\end{split}
\end{equation}
\eqref{ee2} is input-to-state stable (ISS) with respect to $\omega_{i}$ since $A_{ci}$ is Hurwitz. Because in Theorem 2, we have proved that $e_{\sigma(t)}(t)$ is bounded for $t\geq 0$, then $||\omega_{i}||_{\infty}$ is bounded if $||x_{i-1}||_{\infty}$ is bounded. Since system \eqref{ee2} is ISS with respect to $\omega_{i}$, we conclude the boundedness of $||x_{i-1}||_{\infty}$ implies the boundedness of $||x_{i}||_{\infty}$ for $i\in S_{m}$. From the fact that $||x_{0}||_{\infty}$ is bounded, we conclude the boundedness of $||x_{i}||_{\infty}$ for $i\in S_{m}$ \cite{sontag2008input}. 
\section{Numerical Simulations}\label{sim}
Here, we provide numerical simulations to illustrate the
performance of our methods.

\textbf{Example 1.} We consider the $i$-th vehicle in a platoon is using 3 LiDARs with different wavelengths for measuring $d_{i}$ and $d_{i}(t)=5+\sin(t)$. Assume measurement noise in each LiDAR satisfy $\nu_{1}\sim\mathcal{U}(-b_{1},b_{1})$, $\nu_{2}\sim\mathcal{U}(-b_{2},b_{2})$, $\nu_{3}\sim\mathcal{U}(-b_{3},b_{3})$ with $b_{1}=0.1$, $b_{2}=0.2$, $b_{3}=0.3$, and we assume $b_{1}$, $b_{2}$, $b_{3}$ are unknown to us. The attacker has 1 radio signal generator capable of generating pulses of various frequencies to spoof all the 3 LiDARs at different time, and at each time $t$, a pulse of a specific frequency is generated to spoof one of the corresponding LiDAR sensors with the same frequency. Denote the attacked LiDAR at time $t$ as $i_{a}(t)\in\left\lbrace 1, 2, 3\right\rbrace $ and let $\eta_{i_{a}(t)}\sim\mathcal{N}(0,5^{2})$. For $t\in[1,20]$, the vehicle uses \eqref{es1}-\eqref{es3} for fusing the 3 measurements of $d_{i}(t)$. The performance of the fusion algorithm is shown in Figure \ref{fig:fs}.
\begin{figure}[t]\centering
	\includegraphics[width=0.5\textwidth]{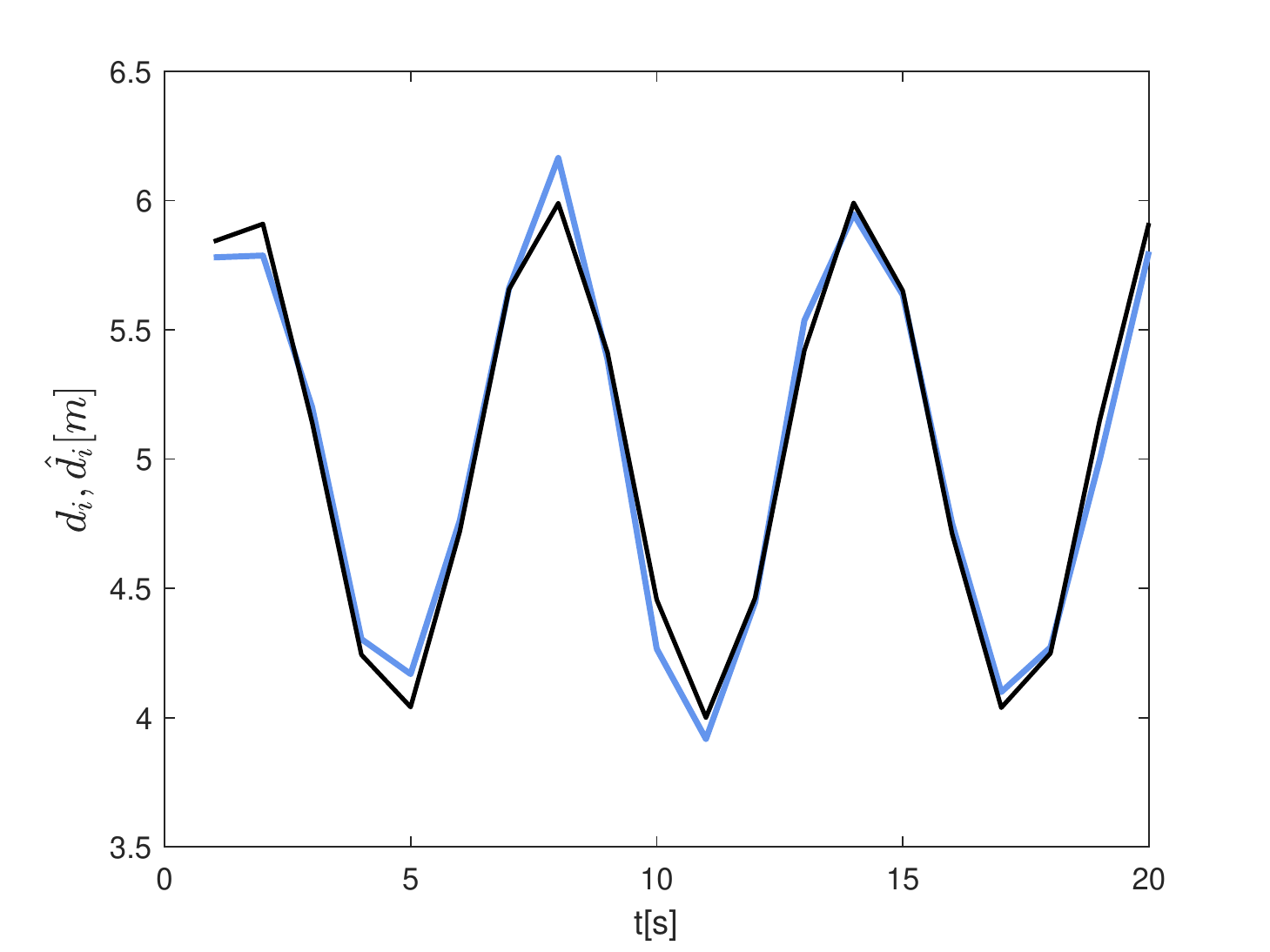}
	\caption{$d_{i}$ (black) and $\hat{d}_{i}$ (blue).}
	\centering
	\label{fig:fs}
\end{figure}

\textbf{Example 2.} We consider the $i$-th vehicle in a platoon is using a LiDAR, a millimeter wave radar and an ultrasonic sensor for measuring $d_{i}$. THe corresponding measurement noise is $\nu_{1}\sim\mathcal{U}(-b_{1},b_{1})$, $\nu_{2}\sim\mathcal{U}(-b_{2},b_{2})$, $\nu_{3}\sim\mathcal{U}(-b_{3},b_{3})$ with $b_{1}=0.1$, $b_{2}=0.4$, $b_{3}=0.5$. The values of $b_{1}$, $b_{2}$ and $b_{3}$ are now assumed to be known. Assume the attacker has one transceiver to generate a pulse to spoof the ultrasonic sensor. Then, the unknown set of attacked sensors $W(k)=\left\lbrace 3\right\rbrace $ is a constant set and let $\eta_{3}(t)\sim\mathcal{N}(0,10^{2})$. Then, $||\nu||_{\infty}=0.5$, and $\tau_{d1}=0.6$, $\tau_{d2}=0.9$, $\tau_{d3}=1$ accordingly. For $t\in[1,1000]$, we run Algorithm \ref{alg:the_alg2} and the result shows that we have $100\%$ successful detection. For $t\in[1,20]$, \eqref{i1}-\eqref{i3} are used for isolating the attacked sensors. The isolation algorithm claims sensor 0 is under attack at time $t$ if no attacked sensors are isolated. The performance of our isolation algorithm is presented in Figure \ref{fig:ff}, where it is shown that our algorithm  successfully isolates the ultrasonic sensor as the attacked one $13$ out of $20$ tests.
\begin{figure}[t]\centering
	\includegraphics[width=0.5\textwidth]{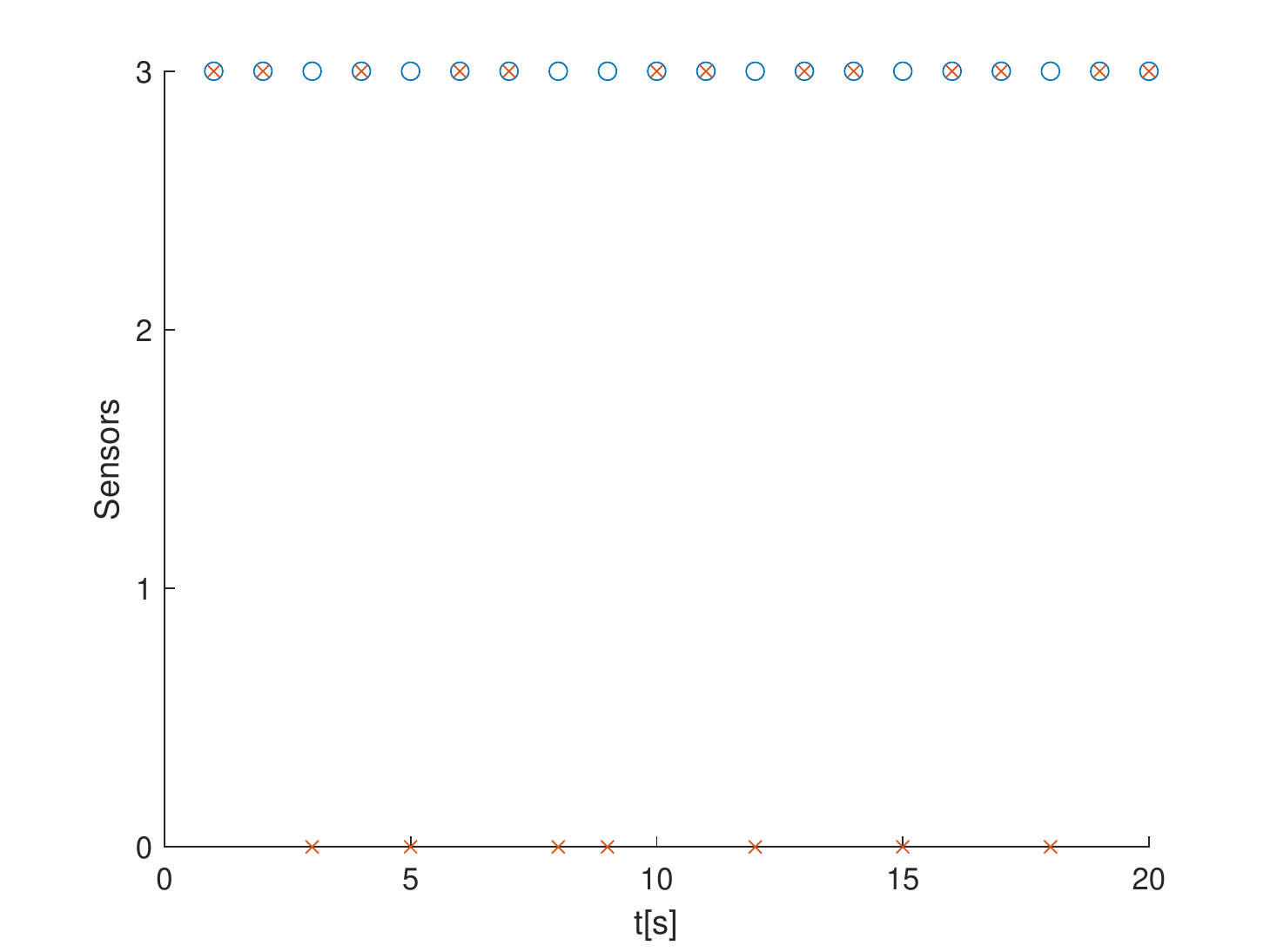}
	\caption{The actual attacked sensor ('o') and the isolated sensor ('x').}
	\centering
	\label{fig:ff}
\end{figure}
\end{section}

\begin{figure}[h]\centering
	\includegraphics[width=0.5\textwidth]{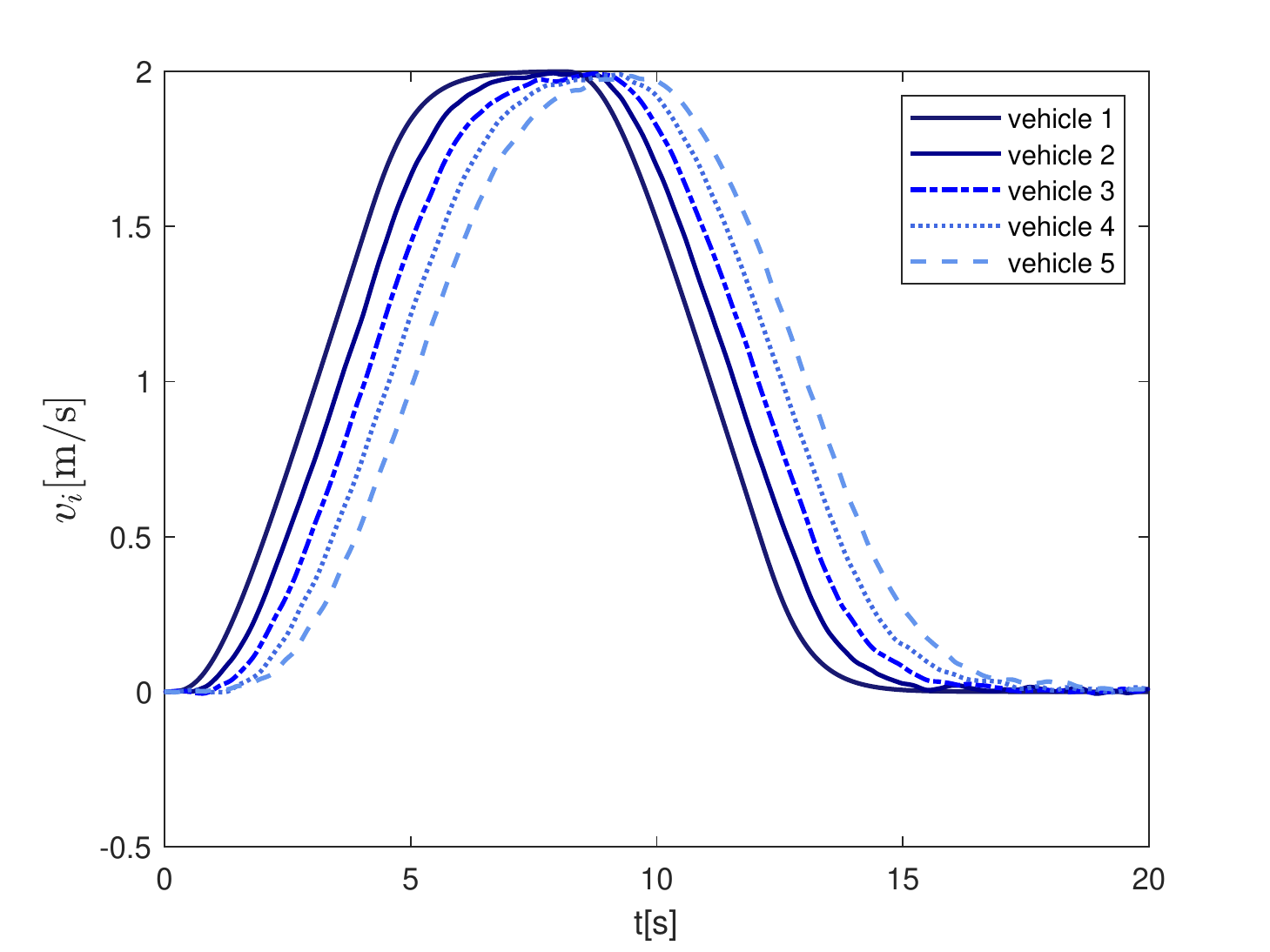}
	\caption{$H_{\infty}$ controller, measured velocity response at startup.}
	\centering
	\label{fig:f4}
\end{figure}

\begin{figure}[h]\centering
	\includegraphics[width=0.5\textwidth]{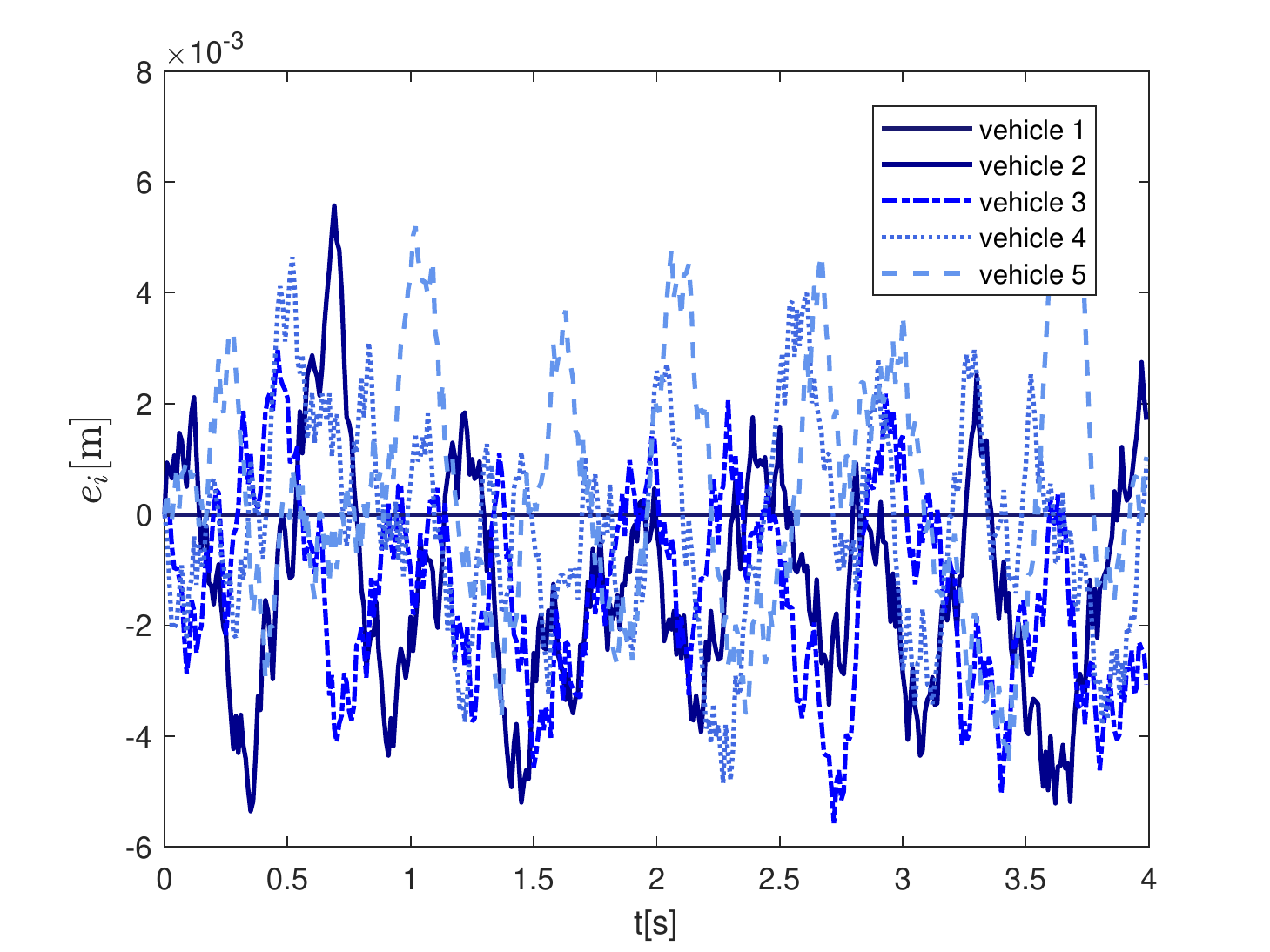}
	\caption{$H_{\infty}$ controller, tracking error at startup.}
	\centering
	\label{fig:f5}
\end{figure}

\textbf{Example 3:} Consider a homogeneous vehicle platoon consisting of $5$ vehicles, where the $i$-th vehicle is equipped with $3$ sensors for measuring $d_{i}$ for all $i\in\left\lbrace 2,3,4,5\right\rbrace $. We let $h_{i}=0.5$, $\tau_{i}=0.1$ for all $i\in\left\lbrace 1,2,3,4,5\right\rbrace $. $\nu_{1}\sim\mathcal{U}(-0.2,0.2)$, $\nu_{2}\sim\mathcal{U}(-0.4,0.4)$, $\nu_{3}\sim\mathcal{U}(-0.6,0.6)$, $\omega_{v},\omega_{u}\sim\mathcal{U}(-0.1,0.1)$, $\xi_{0}=10m/s^{2}$, $0$, $-10m/s^{2}$, $0$, in the time intervals $[0,5]$, $(5,10]$, $(10,15]$, $(15,20]$ seconds respectively. Assume at each time $t\geq 0$, one of the $3$ sensors of each vehicle is randomly selected to be attacked. Denote the attacked sensor of vehicle $i$ at time $t$ as $i_{a}(t)\in\left\lbrace 1, 2, 3\right\rbrace $ and let $\eta_{i_{a}(t)}\sim\mathcal{N}(0,5^{2})$.
For all $t\geq 0$, the $i$-th vehicle $i=2,3,4,5$ uses \eqref{es1}-\eqref{es3} for fusing measurements of $d_{i}(t)$.
We use algorithms 3 and 4 in \cite{He2006} to design a robust controller for each vehicle with $H_{\infty}$ gain $\gamma=1.5235$ and $k_{p}=0.8700$, $k_{d}=11.1683$, $k_{dd}=0.0009$. The performance of the robust controller is shown in Figures \ref{fig:f4}-\ref{fig:f5}.
\section{Conclusion}\label{conclusion}
Exploiting sensor redundancy, we have proposed an attack-resilient sensor fusion framework for connected and automated vehicles. We prove that the proposed sensor fusion algorithm provides a robust estimation of the correct sensor information provided that less than half of the sensors are corrupted by attack signals. The estimation is then used to detect and isolate attacks and stabilize the closed-loop dynamics of each vehicle. We have provided an $H_{\infty}$ controller for each CACC-equipped CAV in the platoon to guarantee its closed-loop stability with reduced joint effects of estimation error, sensor and communication channel noise on platooning string-stable behavior and tracking performance of each vehicle. Our sensor fusion algorithm can be applied to a large class of security-critical cyber-physical systems.
\bibliographystyle{ieeetr}
\bibliography{Observer1}
\end{document}